\documentclass[%
reprint,
english,
 amsmath,amssymb,
 aps,
 prl,
]{revtex4-1}
\usepackage[utf8]{inputenc}
\usepackage[T1]{fontenc}
\usepackage{graphicx}
\usepackage{dcolumn}
\usepackage{bm}
\usepackage[hidelinks]{hyperref}
\usepackage{mathtools}

\begin{document}


\title{Lasing in the Space Charge-Limited Current Regime}

\author{Alex J. Grede}
\email{agrede@ieee.org}
\author{Noel C. Giebink}%
\affiliation{Department of Electrical Engineering, Pennsylvania State University, University Park, Pennsylvania 16802}

\date{\today}

\begin{abstract}
  We introduce an analytical model for ideal organic laser diodes based on the argument that their intrinsic active layers necessitate operation in the bipolar space charge-limited current regime. Expressions for the threshold current and voltage agree well with drift-diffusion modeling of complete \emph{p-i-n} devices and an analytical bound is established for laser operation in the presence of annihilation and excited-state absorption losses. These results establish a foundation for the development of organic laser diode technology.
\end{abstract}

\maketitle

Epitaxial inorganic semiconductors are the basis for nearly all laser diodes today. Organic semiconductors could provide a new, wavelength-tunable laser diode platform, but are disadvantaged by lower charge carrier mobility, lower thermal conductivity, and more efficient excited state quenching interactions. Beyond their material properties, however, organic and inorganic laser diodes also fundamentally differ in their mode of electrical operation: gain in the inorganic case is driven by diffusion current in a \emph{p-n} junction,\cite{coldren_diode_2012}, whereas in the organic case, the need for an intrinsic active layer (since electrical doping severely quenches organic semiconductor excited states\cite{forrest_organic_2020,kohler_electronic_2015}) requires gain to be achieved by a space charge-limited (SCL) drift current. Although the impact of SCL current is well-appreciated for organic light emitting diodes (OLEDs)\cite{forrest_organic_2020}, its implications for organic laser diode operation have not been explored in detail.

Here, we derive expressions for the threshold voltage and current density of ideal organic laser diodes where gain is provided by a bipolar SCL current in the intrinsic active layer of a \emph{p-i-n} device. The results are extended to treat organic laser operation in the presence of triplet exciton and polaron losses, and are validated using the commercial device simulator \textsc{Setfos}\cite{noauthor_setfos_2020}. As organic laser efforts accelerate following the initial demonstration by \textcite{sandanayaka_indication_2019}, the framework established here should prove useful for guiding future development.

Space charge effects become significant in a semiconductor when the transit time of electrons and holes drifting across it (\(\tau_{\mathrm{tr}}=L^2/\mu_{n,p} (V-V_{\mathrm{bi}})\), where \(L\) is the layer thickness, \(V\) is the applied voltage, \(V_{\mathrm{bi}}\) is the built in potential, and \(\mu_{n,p}\) is the electron or hole mobility) is smaller than the dielectric relaxation time, \(\tau_\mathrm{rlx}=\epsilon/q(\mu_n n_0 + \mu_p p_0)\), set by its dielectric constant (\(\epsilon\)) and equilibrium free charge density (\(n_0\), \(p_0\), with \(q\) representing the elementary charge). In essence, this situation corresponds to injecting more charge into the semiconductor than exists in equilibrium and it is frequently the case in undoped organic semiconductors\cite{forrest_organic_2020,kohler_electronic_2015,pope_electronic_1999}. Organic laser diodes are generally expected to operate in this regime because 1) they operate at high current density where drift dominates diffusion and 2) although their electron and hole transport layers are typically doped to achieve high conductivity\cite{sandanayaka_indication_2019}, their active layer (i.e. gain region) must remain intrinsic to avoid strong exciton quenching with ionized dopant and polaron species. Thus, \(\tau_{\mathrm{tr},n}, \tau_{\mathrm{tr},p} \ll \tau_\mathrm{rlx}\) is well-satisfied and, notwithstanding the limited diffusion of charge carriers from the contacts/transport layers into the active layer\cite{wetzelaer_ohmic_2014}, optical gain must be achieved in the bipolar SCL current regime.

The two-carrier SCL current problem was originally solved by Parmenter and Ruppel for the ideal case of a trap-free insulator\cite{parmenter_two-carrier_1959}. The result exhibits the same \(V^2/L^3\) dependence as the unipolar case, but predicts an enhancement of the overall current density that depends on the recombination rate between electrons and holes. In the following, we use this solution to establish analytical bounds and scaling relationships for an ideal organic laser diode with Ohmic electron and hole injection into the active layer; any limitations on injection would constitute a source of non-ideality.

In organic semiconductors with a large exciton binding energy and exchange splitting between bright (singlet) and dark (triplet) exciton states, the net modal gain can be expressed as:
\begin{equation}
  \label{eq:netModalGain}
  g = \Gamma \sigma_{\mathrm{st}} \left(N_{\mathrm{S}}-N_{\mathrm{tr}}\right)-\alpha\,,
\end{equation}
where \(\Gamma\) is the modal confinement factor,
\(\sigma_{\mathrm{st}}\) is the stimulated emission cross-section,
\(N_{\mathrm{S}}\) is the singlet exciton density, \(N_{\mathrm{tr}}\)
is the transparency density, and \(\alpha\) is the net optical loss due to outcoupling, scattering, or parasitic absorption from other materials or excited states. At threshold, \(g=0\) and thus the threshold singlet density is:
\begin{equation}
  \label{eq:cNsthr}
  N_{\mathrm{th}} =
  N_{\mathrm{tr}}\left(1+\frac{\alpha}{N_{\mathrm{tr}}\Gamma\sigma_{\mathrm{st}}}\right)\,.
\end{equation}
The singlet density is governed by the rate equation:
\begin{equation}
  \label{eq:dNsdt}
  \frac{\mathrm{d}N_{\mathrm{S}}}{\mathrm{d}t} = \chi_{\mathrm{S}} R -
  N_{\mathrm{S}}\left(k_{\mathrm{S}} + k_\mathrm{Q}\right) - g v_\mathrm{g} N_{\mathrm{ph}}\,,
\end{equation}
where \(R\) is the exciton generation rate due to Langevin recombination of electrons and holes (i.e. \(R=\gamma np\), where \(\gamma=q(\mu_n+\mu_p)/\epsilon\))\cite{pope_electronic_1999}, \(\chi_{\mathrm{S}}=1/4\) is the singlet spin fraction\cite{segal_excitonic_2003}, \(k_{\mathrm{S}}\) is the natural singlet decay rate, and \(k_{\mathrm{Q}}\) accounts for any additional quenching processes such as exciton-exciton and exciton-polaron annihilation explored below. The term on the right-hand side of Eq.~(\ref{eq:dNsdt}) accounts for stimulated emission, with \(N_{\mathrm{ph}}\) and \(v_{\mathrm{g}}\) equal to the photon density and group velocity of the lasing mode, respectively.

At threshold, \(g=0\) and thus Eq.~(\ref{eq:dNsdt}) can be solved in steady state to yield the threshold recombination rate:
\begin{equation}
  \label{eq:cRth}
  R_{\mathrm{th}} =
  \frac{
  N_{\mathrm{tr}} k_{\mathrm{S}}
  }{
  \chi_{\mathrm{S}}
  }
  \left(1+\frac{k_{\mathrm{Q}}}{k_{\mathrm{S}}}\right)
  \left(1+\frac{\alpha}{\Gamma \sigma_{\mathrm{st}}N_{\mathrm{tr}}} \right),
\end{equation}
which factors into an ideal rate multiplied by exciton and optical loss terms in parentheses. Equation~(\ref{eq:cRth}) is derived assuming a spatially uniform exciton density, which is rigorously valid for the ideal case of \(\mu_n=\mu_p\) as shown below, and thus also justifies the neglect of exciton diffusion in Eq.~(\ref{eq:dNsdt}).

The Parmenter-Ruppel solution subsequently establishes the link between recombination rate, voltage, and current density in the device. The general result is provided in the Supplementary Material; however, we focus here on the specific case of equal electron and hole mobilities (\(\mu_n = \mu_p = \mu\)) and perfect charge balance (i.e. every injected electron recombines with a hole in the active layer) since it constitutes the limit of an ideal organic laser diode.

\begin{figure}
  \centering
  \includegraphics{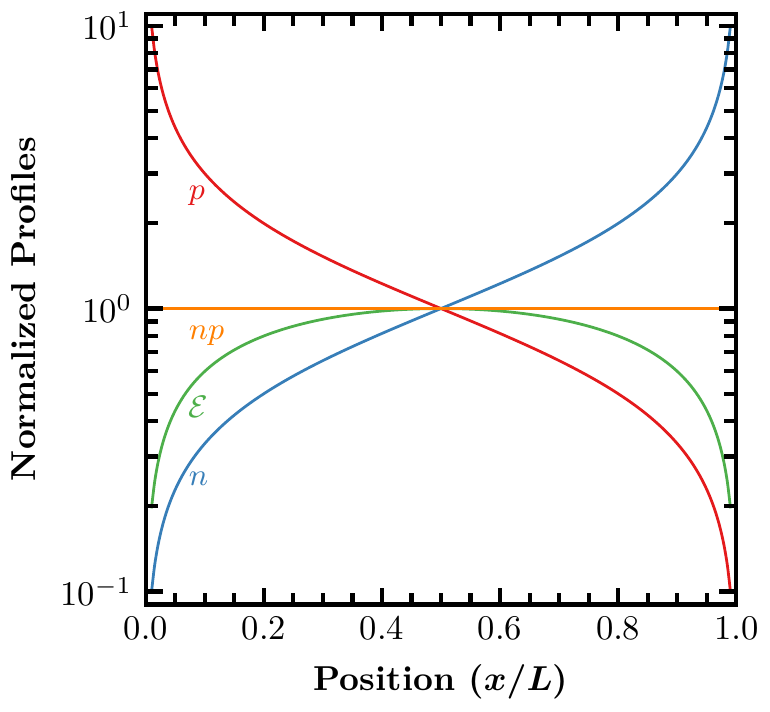}
  \caption{Normalized carrier concentration, electric field, and \(np\) product as a function of position for bipolar space charge-limited current with a Langevin recombination rate, equal charge carrier mobilities, and Ohmic electron and hole injection.}
  \label{fig:carConc}
\end{figure}

The current-voltage relationship is given by the Mott-Gurney law:
\begin{equation}
  \label{eq:childsLaw}
  J = \frac{9}{8} \epsilon \mu_{\mathrm{eff}} \frac{\left(V-V_{\mathrm{bi}}\right)^2}{L^3},
\end{equation}
but with an effective mobility, \(\mu_{\mathrm{eff}} = 256\mu/(9\pi^2)\approx 2.9\mu\), larger than that of the individual carriers due to the neutralization of space charge by recombination. For the assumed case of Ohmic electron and hole injection into the active layer, the built-in potential is approximated by the bandgap energy, \(V_{\mathrm{bi}} \approx E_\mathrm{g}/q\). The position, electric field, and carrier densities are most conveniently expressed parametrically in terms of the fractional electron current \(j_n = J_n / J\):
\begin{subequations}
  \label{eq:lav}
  \begin{alignat}{2}
  x &= L j_n &
    \mathcal{E}
    &=
    \sqrt{j_n \left(1 - j_n\right)} \sqrt{\frac{2 L J}{\epsilon \mu}}\label{eq:lavPosE}\\
    n
    &=
    \sqrt{\frac{j_n}{1-j_n}}
    \sqrt{
      \frac{
        \epsilon J}
      {
        2 q^2 \mu L
      }}
    &\quad
    p
    &=
    \sqrt{\frac{1-j_n}{j_n}}
    \sqrt{
      \frac{
        \epsilon J}
      {
        2 q^2 \mu L
      }} .\label{eq:lavCar}
  \end{alignat}
\end{subequations}
Each quantity scales linearly with the applied voltage (square root of the current) and is shown normalized to its value at the midpoint of the active layer in Fig.~\ref{fig:carConc}. The \(np\) product is spatially uniform as evident from inspection of Eq.~(\ref{eq:lavCar}), justifying the neglect of spatial variation in deriving Eq.~(\ref{eq:cRth}). As in the case of unipolar SCL current, these equations break down in the immediate vicinity of the contacts as the field drops to zero and transport becomes diffusive.

It is subsequently straightforward to evaluate the current density and voltage in terms of the recombination rate at threshold:
\begin{equation}
  \label{eq:lavJVP}
    J_{\mathrm{th}} = q L R_{\mathrm{th}} \quad\mathrm{and}\quad
    V_{\mathrm{th}} = \frac{\pi L^2}{4} \sqrt{\frac{q
                      R_{\mathrm{th}}}{2 \epsilon \mu}}+V_\mathrm{bi} .
\end{equation}
The expression for \(J_\mathrm{th}\) follows directly from particle conservation due to the assumption of perfect charge balance and shows that the simple dimensional analysis estimate commonly used in the literature\cite{kozlov_structures_2000} is exact for a bipolar SCL current. We note that, although Eq.~(\ref{eq:lavJVP}) is derived under the assumption \(\mu_n = \mu_p\), it remains reasonably accurate even when the mobilities differ by up to an order of magnitude provided that their average value is used for \(\mu\); see the Supplementary Material for details.

In evaluating \(R_{\mathrm{th}}\) from Eq.~(\ref{eq:cRth}), we note that, while \(N_{\mathrm{tr}}\) can be obtained from thermodynamic considerations (see the Supplementary Material), it is immaterial in practice because the cold cavity loss (\(\alpha_{\mathrm{cav}}\)) typically satisfies \(\alpha_{\mathrm{cav}} / \Gamma \sigma_{\mathrm{st}} N_{\mathrm{tr}} \gg 1\) and thus \(R_{\mathrm{th}} \approx N_{\mathrm{th},0} k_{\mathrm{S}} / \chi_{\mathrm{S}}\), which is readily determined from the threshold singlet density (\(N_{\mathrm{th},0}\)) measured experimentally under impulsive optical pump conditions. Taking parameter values typical of the 4,4$^{\prime}${}-bis[(N-carbazole)styryl]bi-phenyl (BSBCz) diode lasers reported by \textcite{sandanayaka_indication_2019}: \(N_{\mathrm{th},0}=2\times10^{16}\){}~cm$^{-3}$, \(k_{\mathrm{S}}=1\){}~ns$^{-1}$, \(\epsilon = 4 \epsilon_0\), \(\mu \approx 1\times10^{-3}\){}~cm$^2\,${}V$^{-1}\,${}s$^{-1}${}, \(E_\mathrm{g}\approx2.7\){}~eV, and \(L= 150\){}~nm, we obtain \(J_{\mathrm{th}} = 190\){}~A$\,${}cm$^{-2}${}, \(V_{\mathrm{th}} = 26\){}~V and a threshold power density \(P_{\mathrm{th}} = J_\mathrm{th} V_\mathrm{th} = 5.1\){}~kW$\,${}cm$^{-2}${} as lower bounds for laser operation. For comparison, the experimentally-recorded values are \(J_{\mathrm{th}}=600\){}~A$\,${}cm$^{-2}${}, \(V_{\mathrm{th}} = 34\){}~V and
\(P_{\mathrm{th}} = 20\){}~kW$\,${}cm$^{-2}${}\cite{sandanayaka_indication_2019}.

Equation~(\ref{eq:lavJVP}) highlights the importance of active layer thickness for organic laser diodes. Whereas \(J_{\mathrm{th}}\) depends weakly on \(L\) since \(\Gamma\propto L\) to first order, the threshold voltage and power density both scale as \(L^{3/2}\). The is important in the context of thermal management because organic laser diodes not only generate more heat than their inorganic counterparts (due to their higher voltage), but have more difficulty dissipating it (due to their lower thermal conductivity) and are less able to withstand high temperature without degrading. Electric field strength, which scales as \(L^{1/2}\), is another concern since its maximum in the example above (\(\mathcal{E}_{\mathrm{max}}=2\){}~MV$\,${}cm$^{-1}${}) is comparable to the \(\sim3\){}~MV$\,${}cm$^{-1}${} breakdown field of many organic semiconductors\cite{forrest_organic_2020,kohler_electronic_2015,pope_electronic_1999}. To this point, dielectric breakdown was reported after roughly fifty pulses near threshold in Ref.~\cite{sandanayaka_indication_2019}.

\begin{figure}
  \centering
  \includegraphics{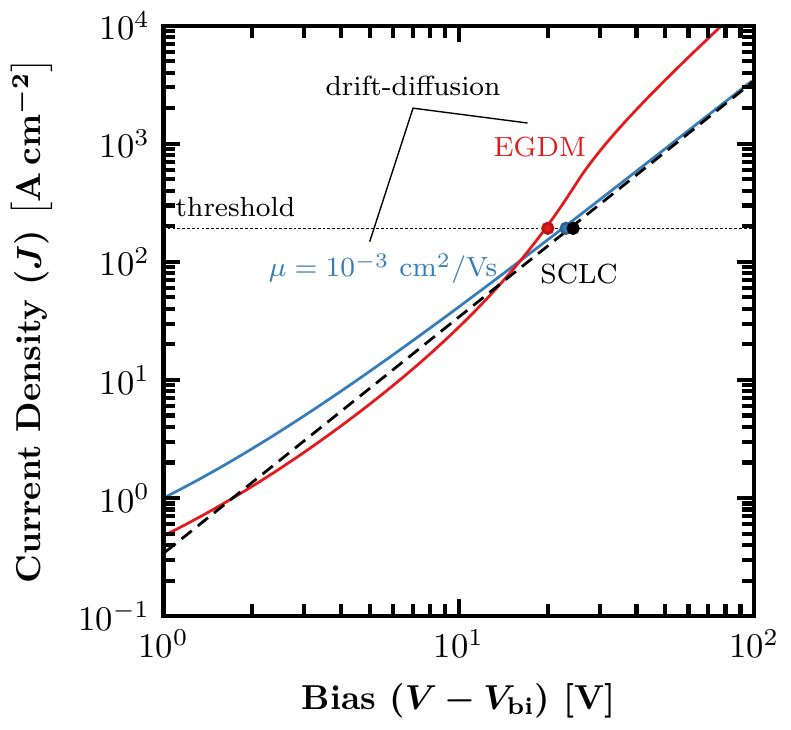}
  \caption{Drift-diffusion simulations of the current-voltage relationship for a BSBCz-like active layer with Ohmic majority carrier injection and equal electron and hole mobilities that are either constant (blue line) or given by the extended Gaussian disorder model (EGDM) for disordered organic semiconductors (red). The black dashed line shows the analytical space charge-limited current (SCLC) prediction and solid circles denote the lasing threshold in each case.}
  \label{fig:JV}
\end{figure}

Given that the threshold relations above neglect the existence of charge transport layers and also assume infinite carrier densities at the active layer interfaces, it is important to test the accuracy of these results against full drift-diffusion modeling of a real device architecture. Retaining the BSBCz parameters from above, we treat the case of an organic laser with a 150~nm-thick intrinsic active layer and fix the majority carrier concentrations at each edge to $2.5\times10^{19}${}~cm$^{-3}$ to simulate Ohmic injection from heavily doped transport layers. Drift-diffusion simulations are carried out using the commercial software \textsc{Setfos}\cite{noauthor_setfos_2020} and the results are presented in Fig.~\ref{fig:JV} for the case of a constant mobility (\(\mu_\mathrm{n}=\mu_\mathrm{p}=1\times10^{-3}\){}~cm$^2\,${}V$^{-1}\,${}s$^{-1}${}; blue) and for the case in which it depends locally on electric field and carrier density according to the extended Gaussian disorder model (EGDM, red); details of the EGDM are provided in the Supplementary Material. The threshold current density in both simulations is in good agreement with the analytical prediction; however, the threshold voltage in the EGDM case is slightly lower. This is due to the increase in mobility with field and carrier concentration in the EGDM model, which highlights the importance of using mobility values that are congruent with the conditions at threshold.

From Fig.~\ref{fig:carConc}, it is evident that to sustain the electron and hole concentrations in the middle of the device (and thus the \(np\) product everywhere), the majority carrier concentrations injected at the active layer edges must be roughly an order of magnitude higher. Inserting \(J_{\mathrm{th}} = 190\){}~A$\,${}cm$^{-2}${} from Fig.~\ref{fig:JV} into Eq.~(\ref{eq:lavCar}) suggests that edge carrier concentrations as low as $\sim3\times10^{18}${}~cm$^{-3}$ are sufficient to maintain charge balance up to the lasing threshold, in agreement with the full drift-diffusion model. This is important because it sets the minimum doping concentration required for the transport layers in a laser diode (i.e. due to continuity of \(n\) and \(p\)) which, notably, is within the range typical for \emph{p-i-n} OLEDs\,\cite{walzer_highly_2007}; more detail on this point is provided in the Supplementary Material.

At this stage, it is also important to assess the impact of triplet exciton and polaron-related optical losses (due to respective absorption cross-sections \(\sigma_\mathrm{TT}\)\,\cite{zhang_existence_2011} and \(\sigma_\mathrm{PP}\) at the lasing wavelength) and quenching interactions (with respective annihilation rate coefficients \(k_\mathrm{STA}\) and \(k_\mathrm{SPA}\))\,\cite{baldo_prospects_2002,gartner_influence_2007}. Because the latter typically depend on the former through F\"{o}rster energy transfer\cite{yokota_effects_1967}, absorption and annihilation losses must be treated on equal footing. Assuming only one species of polaron (holes in this case) is detrimental, the quenching rate in Eq.~(\ref{eq:dNsdt}) becomes \(k_{\mathrm{Q}} = k_{\mathrm{SPA}}p_\mathrm{mid}+k_{\mathrm{STA}} N_{\mathrm{T}}\) and the loss in Eq.~(\ref{eq:netModalGain}) becomes \(\alpha = \alpha_{\mathrm{cav}}+\Gamma(\tfrac{\pi}{2} p_\mathrm{mid} \sigma_{\mathrm{PP}}+N_{\mathrm{T}} \sigma_{\mathrm{TT}})\), where \(p_\mathrm{mid}\) is the hole density at the midpoint of the active layer. Using the threshold singlet density in the absence of triplet and polaron losses, \(N_{\mathrm{th},0}\) (defined in Eq.~(\ref{eq:cNsthr}) by the cold cavity loss), Eq.~(\ref{eq:cRth}) can be rewritten in the form of an implicit quartic equation:
\begin{widetext}
\begin{equation}
    \label{eq:RthAA}
    R_{\mathrm{th}} = \frac{N_{\mathrm{th},0} k_{\mathrm{S}}}{\chi_{\mathrm{S}}}
    \left(
    1+
    \frac{R_{\mathrm{th}} \chi_{\mathrm{T}} t_{\mathrm{rise}} k_{\mathrm{STA}}}{k_{\mathrm{S}}}+
    \frac{k_{\mathrm{SPA}}}{k_{\mathrm{S}}}
    \sqrt{\frac{\epsilon R_{\mathrm{th}}}{2q\mu}}\right)
    \left(
    1+
    \frac{
    R_{\mathrm{th}} \chi_{\mathrm{T}} t_{\mathrm{rise}} \sigma_{\mathrm{TT}}
    }{
    N_{\mathrm{th},0}\sigma_{\mathrm{st}}
    }+
    \frac{
    \pi \sigma_{\mathrm{PP}}
    }{
    2 N_{\mathrm{th},0}\sigma_{\mathrm{st}}
    }
    \sqrt{\frac{\epsilon R_{\mathrm{th}}}{2 q \mu}}\right).
\end{equation}
\end{widetext}
In this expression, the rise time (\(t_\mathrm{rise}\)) is important because it determines the extent to which long-lived triplet excitons accumulate before the full current density is achieved (i.e. \(N_{\mathrm{T}} \approx R \chi_{\mathrm{T}} t_{\mathrm{rise}}\) by the time the full recombination rate, \(R\), is reached)\cite{giebink_temporal_2009}. Although \(t_\mathrm{rise}\) is nominally characteristic of the electrical pulse, it cannot be significantly faster than the transit time of charge carriers drifting across the active layer since this is the time it takes to establish the SCL recombination profile to begin with (\(\tau_\mathrm{tr}\approx10\){}~ns for the BSBCz example above)\,\cite{many_theory_1962}. Note also that, in seeking to approximate the spatially nonuniform polaron density with a single effective value for annihilation and absorption, it is more accurate to use the midpoint density for the former and the layer-averaged density (\(\tfrac{\pi}{2} p_\mathrm{mid}\)) for the latter as discussed in the Supplementary Material.

\begin{figure}[b]
  \centering
  \includegraphics{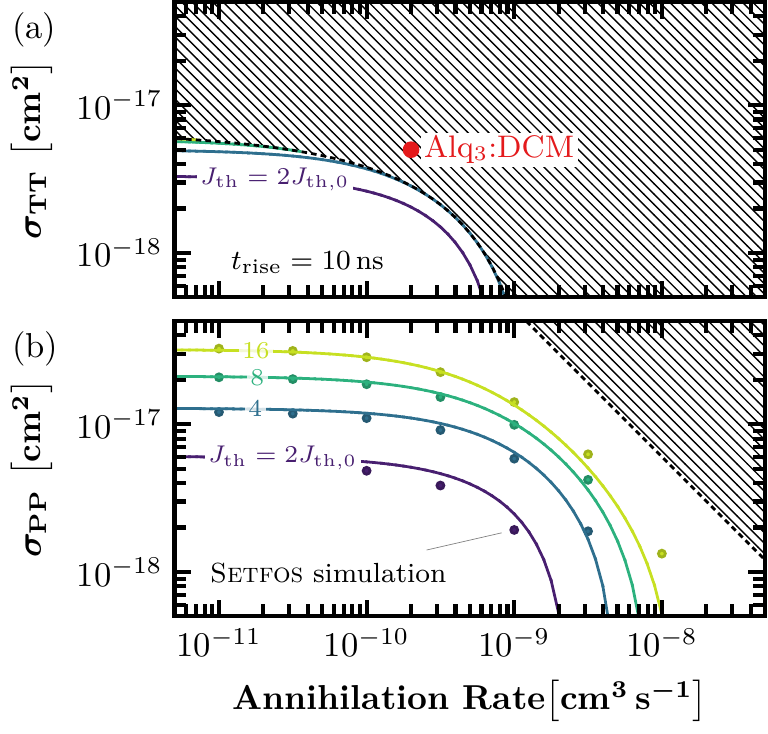}
  \caption{Contours showing the relative increase in threshold current due to  parasitic absorption and singlet exciton annihilation caused by (a) triplet excitons and (b) hole polarons. Lasing is forbidden in the shaded region at any current density. The solid lines are calculated using Eq.~(\ref{eq:RthAA}) and are compared in (b) with the results of \textsc{Setfos} numerical modeling (solid circles). The Alq\(_3\):DCM material system marked by the red dot in (a) lies in the forbidden region for the \(t_\mathrm{rise}=10\){}~ns case shown. All of the calculations are based on the BSBCz parameters given in the main text along with \(\sigma_{\mathrm{st}}=2\times10^{-16}\){}~cm$^2${}\cite{nakanotani_spectrally_2007}.}
  \label{fig:Quenching}
\end{figure}

The form of Eq.~(\ref{eq:RthAA}) is useful because it allows the impact of each loss mechanism to be understood individually. For example, singlet-triplet annihilation on its own doubles the threshold when the middle term in the first set of parentheses is equal to unity, which corresponds to the situation where \(t_\mathrm{rise}=\chi_\mathrm{S}/(2\chi_\mathrm{T}N_{\mathrm{th},0}k_{\mathrm{STA}})\). More generally, for Eq.~(\ref{eq:RthAA}) to have \emph{any} physically meaningful solution, the absorption and annihilation coefficients must satisfy the following inequalities:
\begin{subequations}
\label{eq:RthAAin}
\begin{gather}
    \sqrt{\sigma_{\mathrm{TT}}/\sigma_{\mathrm{st}}} + \sqrt{N_{\mathrm{th},0}k_{\mathrm{STA}}/k_{\mathrm{S}}} <
    \sqrt{\chi_{\mathrm{S}}/\left(\chi_{\mathrm{T}} k_{\mathrm{S}} t_{\mathrm{rise}}\right)} \label{eq:RthAAinTrip}\\ \mathrm{and} \quad
    k_{\mathrm{SPA}} \sigma_{\mathrm{PP}} < 4 \chi_{\mathrm{S}} \sigma_{\mathrm{st}} q \mu / \left( \pi \epsilon \right) \label{eq:RthAAinPol}.
\end{gather}
\end{subequations}
Figures~\ref{fig:Quenching}(a) and 3(b) plot these bounds for triplet and polaron losses, respectively, along with contours that show the relative increase in threshold caused by each species. Figure~\ref{fig:Quenching}(b) also includes the results from full \textsc{Setfos} numerical modeling (solid symbols) of the polaron case, validating the midpoint and average carrier density approximations used to derive Eq.~(\ref{eq:RthAA}).

These results highlight the difference between triplet losses, where a small increase in absorption cross-section or annihilation coefficient can mean the difference between a modest threshold increase and prevention of lasing outright, and polaron losses, where the penalty is more gradual and does not depend on external factors like \(\alpha_\mathrm{cav}\) or \(t_\mathrm{rise}\). A blunt way of characterizing this difference is that triplet excitons are either insignificant or catastrophic, whereas the region of parameter space between these extremes is broader for polarons. This highlights the importance of designing new organic gain media like BSBCz that have low overlap between their emission and triplet-triplet absorption  spectra \cite{sandanayaka_indication_2019}, and of implementing device architectures that can deliver high-speed electrical pulses\cite{chime_electrical_2018}, since this is the difference between success and failure for the classic Alq\(_3\):DCM\cite{giebink_temporal_2009} gain medium in Fig.~\ref{fig:Quenching}(a).

Another intuitive, but important guideline is to seek organic gain media with equal (and maximal) electron and hole mobilities. To the extent that perfect charge balance is maintained and the envelope of the lasing mode varies negligibly over the active layer, the general solution for \(\mu_n \ne \mu_p\) given in the Supplementary Material shows that Eq.~(\ref{eq:lavJVP}) is largely unaffected when the mobility of one carrier dominates (in this case \(\mu_{\mathrm{eff}}\) is just the mobility of the more mobile charge carrier). The problem with imbalanced mobility, however, is that it concentrates recombination toward the lower mobility carrier side of the active layer. This not only exacerbates annihilation loss, but also makes it more challenging to maintain charge balance since the less mobile carrier must be injected at higher density to sustain the same total recombination rate in a narrower region of space; full details are provided in the Supplementary Material.

Finally, we note that the framework developed here may also be relevant for metal halide perovskite (MHP) lasers. Though SCL current is less well-studied in this material class (due in part to complications with ion movement)\cite{duijnstee_toward_2020,sajedi_alvar_space-charge-limited_2020}, the observation that MHP light emitting diodes use undoped active layers\cite{gunnarsson_electrically_2020,liu_metal_2020} along with the fact that electrical doping is a basic challenge for these materials to begin with\cite{duijnstee_toward_2020} makes it plausible that a future MHP laser diode will operate in the bipolar SCL current regime. In this case, however, recombination tends to be sub-Langevin (i.e. the bimolecular recombination coefficient is substantially smaller than the Langevin rate), which causes bipolar SCL current to take place in the injected plasma regime where \(n \approx p\) everywhere\cite{lampert_volume-controlled_1961}. Unfortunately, the analytical solution in this limit can greatly overestimate the threshold current because diffusion strongly modifies the \(np\) product (and thus the recombination current) near the active layer edges (see the Supplementary Material for an example). A full drift-diffusion model is therefore required to accurately describe MHP lasers.

In conclusion, we have put forth a model for organic laser diodes that operate in the bipolar space charge-limited current regime. We have obtained analytical expressions for the threshold voltage and current density and have identified fundamental limits for laser operation in the presence of parasitic annihilation and excited state absorption losses. These results, together with the experiments of \textcite{sandanayaka_indication_2019}, emphasize a shift in strategy from OLED-like architectures characterized by many heterojunctions, blocking layers, and so forth, to a \emph{p-i-n} structure based on a single, low threshold material that can be degenerately \emph{p-} and \emph{n-}doped, and that has no triplet or polaron absorption overlapping with its emission. With rational design of organic laser materials now emerging to meet these criteria\,\cite{ou_computational_2020} and the model here to predict device performance, the future of organic laser diode technology is coming into clearer focus.

\begin{acknowledgments}This work was supported in part by AFOSR Award no. FA9550-18-1-0037 and DARPA Award no N66001-20-1-4052.
\end{acknowledgments}


%

\end{document}